\documentclass[onecolumn]{article}
\usepackage{graphics}
\usepackage[square,comma,longnamesfirst,numbers,sort&compress]{natbib}
\usepackage[pdftex]{graphicx}
\usepackage{color}

\usepackage{amssymb,amsfonts,amsmath}
\usepackage[nomarkers]{endfloat}

\begin{document}

\title{Numerical modeling of non-woven fiber mats: \\Their effective mechanical and electrical properties}

\author{Enis Tuncer$^1$ and Roy l'Abee$^2$\\ {\small $^1$Dielectrics \& Electrophysics Lab, General Electric Global Research Center}\\ {\small Niskayuna NY 12309}\\ {\small $^2$SABIC, Bergen op Zoom}\\ {\small The Netherlands ADC E174}}

\maketitle
\begin{abstract}
  Numerical simulations on non-woven-fibrous, porous structures were performed to determine material design space for energy storage device (battery and ultracapacitor) separators. Material simulations were performed initially with a commercial program called GeoDict using its demo version. Later, in-house computational tools were developed and employed. The numerical routines were created to model mechanical and electrical properties of porous structures. The tools were built as a pre-processor for a commercial finite element package. Effective properties were estimated in the post-processing phase using the current and stress distributions. No multi-physics assumptions were considered to couple electrical and mechanical fields at this stage. The numerical results between two numerical platforms, GeoDict and in-house tools. 
Regions of interest in porosity for battery separators discussed.

| non-woven mats | porous materials | effective properties of materials 
\end{abstract}

\section{Introduction}
Porous structures are used as separators or membranes for battery and ultracapacitor applications \cite{Arora2004,Zhang2007351}. The separator is one of the  critical components in these novel energy storage units. It prevents contacting of the positive and the negative electrodes while providing pathways for ionic charge transport through the ionically conductive electrolyte solution without electonic conduction. The separator material should be chemically reistant to the electrolyte solvents and the reducing and oxidizing environments during cell charging and discharging. It should have mechanical integrity to withstand the high tension during the assembly of the energy storage units. Battery performance can be improved by finding the optimum balance between low fiber volume fraction (facilitates ionic transport) yet a high enough fiber volume fraction to assure sufficient mechanical integrity. 

The separators are classified by \citet{Zhang2007351} (i) microporous polymer membranes, (ii) non-woven fabric mats and (iii) inorganic composite membranes. Each class has its advantages. For example as listed by \citet{Zhang2007351} the microporous polymer membranes are thin and have thermal shutdown capability; the non-woven mats are porous and economical compared to others; finally the composite membranes wet easily with electrolyte and operate at high temperatures. Low temperature capable and inorganic separators find applications with better safety in Li-ion batteries. Low temperature polymers as polyethylene (PE) and polypropylene (PP) structures and  PE and PP blends operate as a fuse by allowing meltdown of the low-temperature-melting-point PE that shuts down the charge conduction backbone--where the internal structure is altered with closed cells and the new porosity destroys percolating paths for the charge carriers--without causing overheating and thermal runway. The ceramic composite separators on the other hand provide high stability and keep dimensional tolerances at high temperatures providing their function--to separate the electrodes in high temperature operation. \citet{Arora2004} have published an excellent review on the materials and properties of separators.

The current study concentrates on the modeling of polymeric separators. 
It investigates mechanical and electrical properties through effective medium theories with implementing combination of numerical routines. Since the fiber spinning is a random process, the numerical simulations need to be adopted to fully capture the randomness in structures. For this purpose, pre- and post-processing tools were developed to be used with a commercial finite element software \cite{comsol43}. Our aim is to determine material manufacturing space for best performance.

\section{Structure of materials simulated}
\begin{figure}[[ht]
  \centering
  \includegraphics[width=.45\linewidth]{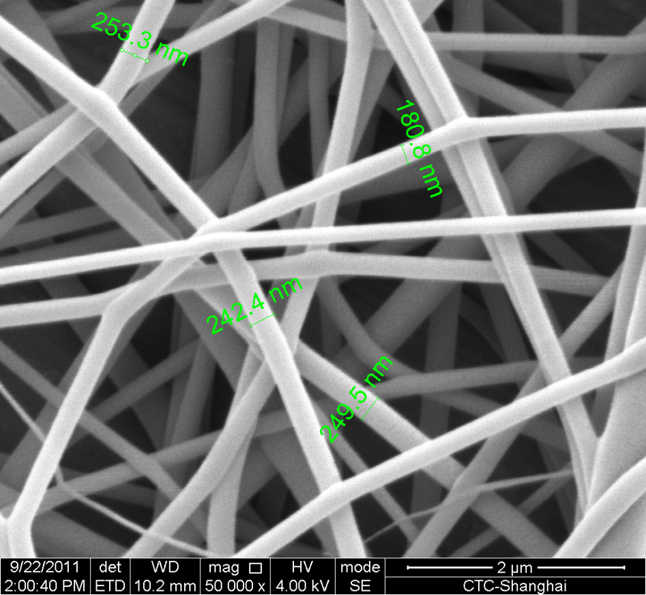}
  \includegraphics[width=.45\linewidth]{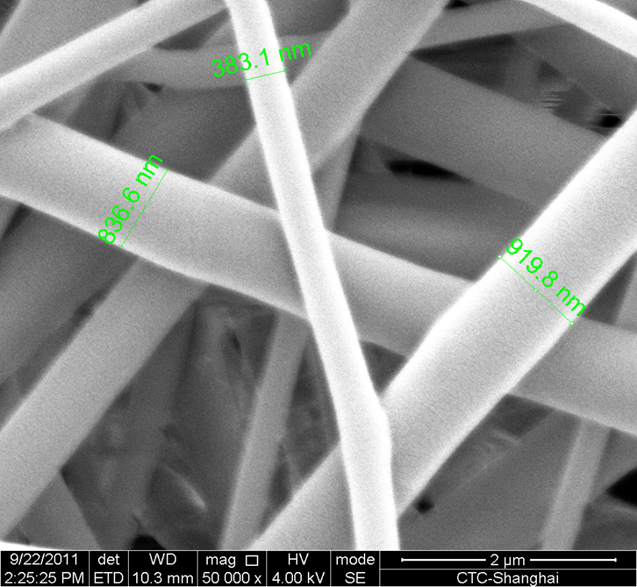}
  \caption{Scanning electron microscope images of electrospan polymer. 
The dimensions of the fibers are shown inside the images.}
  \label{fig:semPBT}
\end{figure}

Material structures simulated in the investigations were 
produced by electrospinning method \cite{Huang2003}. Example of structures from scanning electron microscopy analysis are shown in Fig.~\ref{fig:semPBT}. Observe that the fibers are distributed randomly and have polydispersed dimensions. All the fibers lay on a lateral direction on a substrate--no fibers are perpendicular to the substrate orientation. 

Some of the factors effecting the structure of the fibrous membrane (separator) can be listed\cite{Huang2003} as follows; (1) electrospinning needle thickness and its thickness uniformity; (2) moisture level in the chamber; (3) spinneret  size; (4) solution concentration; (5) solution feeding rate, pressure at the nozzle; (6) applied electric field at the nozzle; (7) temperature; (8) post-treatment with temperature and pressure. One can optimize the fibrous material structure with adjusting these parameters. Post-treatment methods of non-woven mats are also possible to improve/change structure and porosity. Designing experiments and analyzing the fabricated structures are time consuming efforts based on Edisonian-approach. One solution could be to perform numerical simulations with structures close to reality, that can be used to screen structures and establish critical parameters for manufacturing. Here we are attempting to do this with the Monte Carlo based finite element method.  

\section{Modeling background}

Due to increase in need for funtional materials for mechanical, thermal and electrical applications, fibrous structures have become important because of desired properties at low percolation thresholds--one can design anisotropic materials with little addition of fibrous components. Building models on these systems is an area that needs special attention. It is well know that the internal structure of a composite influences the effective properties of the composite \cite{Tuncer2D,TuncerPhysD,Tuncer2002a,TuncerEyeTruss2003,Altendorf2011,Zamel2010}. However, one needs to perform numerous calculations to determine properties of random structures and understand the influence of various parameters on the properties. It is therefore of importance to perform the Monte Carlo based studies. 

In the current study a softcore approach was adapted as discussed by \citet{Altendorf2011}. Similar to their approach infinite lines were generated. Later the lines were converted to a three dimensional cylinder using a moving spherical representational volume. The line was used as the center of the sphere. The volume was included to a large matrix of the modeling domain (cube) which was considered to be $100\times100\times100$. Since we were interested in the effective properties the actual physical dimensions of the cube is irrelevant. Such studies would be important when the physical laws would be different at different scales and high frequency effects due to scattering of acoustic (pressure) and electromagnetic waves are of important.

\begin{figure}[ht]
  \centering
  \includegraphics[width=.5\linewidth]{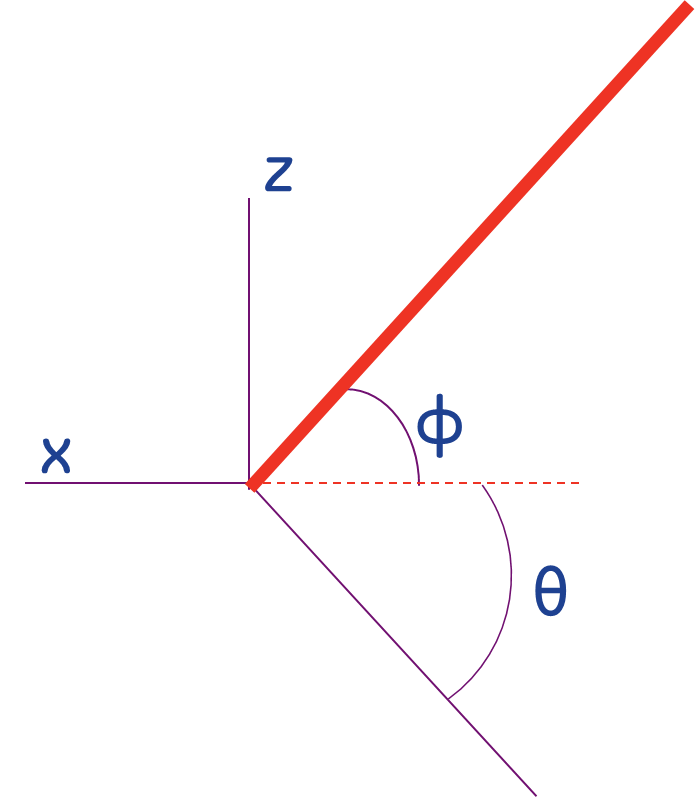}
  \caption{Geometrical definition of the lines used to generate fibers.}
  \label{fig:line}
\end{figure}

\begin{figure}[ht]
  \centering
  \includegraphics[width=.5\linewidth]{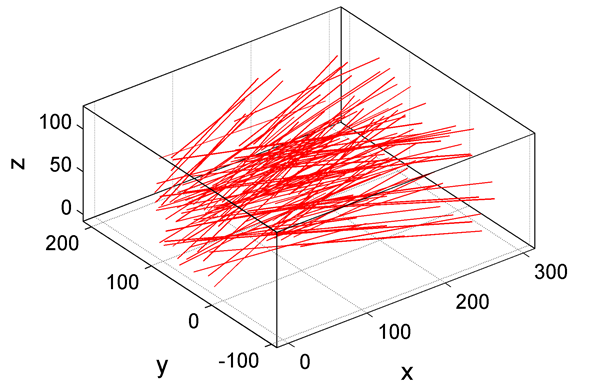}
  \caption{Composition of many lines before generating softcore structures.}
  \label{fig:manylines}
\end{figure}

\begin{figure}[ht]
  \centering
  \includegraphics[width=.5\linewidth]{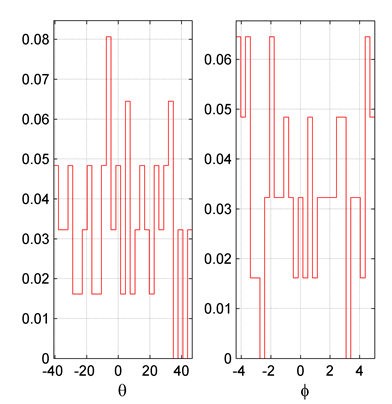}  
  \caption{Fiber distribution as a function of angle.}
  \label{fig:distangle}
\end{figure}

\begin{figure}[ht]
  \centering
  \includegraphics[width=.5\linewidth]{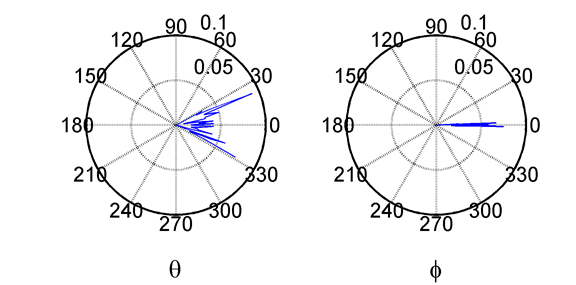}  
  \caption{Fiber distribution as a function of angle in polar representation.}
  \label{fig:distpolar}
\end{figure}

The lines were generated using the definitions in Fig.~\ref{fig:line}. The direction of the line was determined from a random angle distribution where $\phi$ and $\theta$ were adjusted for each line. Each fiber was considered to be started on a plane and extended in the two perpendicular directions in the space.
An example of the distribution of lines for a generated structure is shown in Fig.~\ref{fig:manylines}. 
For this non-woven fiber structure in Fig.~\ref{fig:manylines}, the distribution of angles are shown in Fig.~\ref{fig:distangle} and \ref{fig:distpolar}, where uniform distribution of angles were considered with in a range of values. The polar representation in Fig.~\ref{fig:distpolar} shows a clear picture of the line alignments. Observe that the lines are on a lateral surface with $\phi\sim0$. Each line later converted to a cylinder. More sophisticated approches have been adopted to generate non-woven mats in the literature \cite{Altendorf2011,Zamel2010}

\begin{figure}[ht]
  \centering
  \includegraphics[width=.5\linewidth]{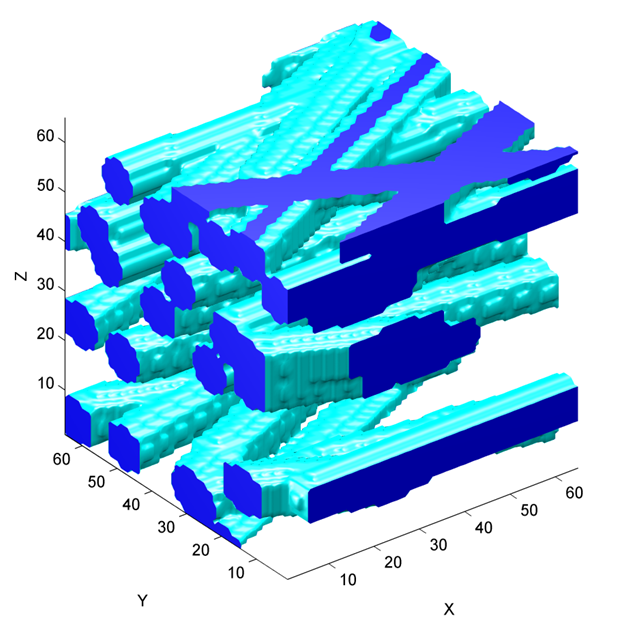}  
  \caption{Section of the fibrous structure after convering lines to three dimensional solid. The size of the cube can be adjusted depending on the finite size scaling.}
  \label{fig:threeDmatlab}
\end{figure}

\begin{figure}[ht]
  \centering
  \includegraphics[width=.5\linewidth]{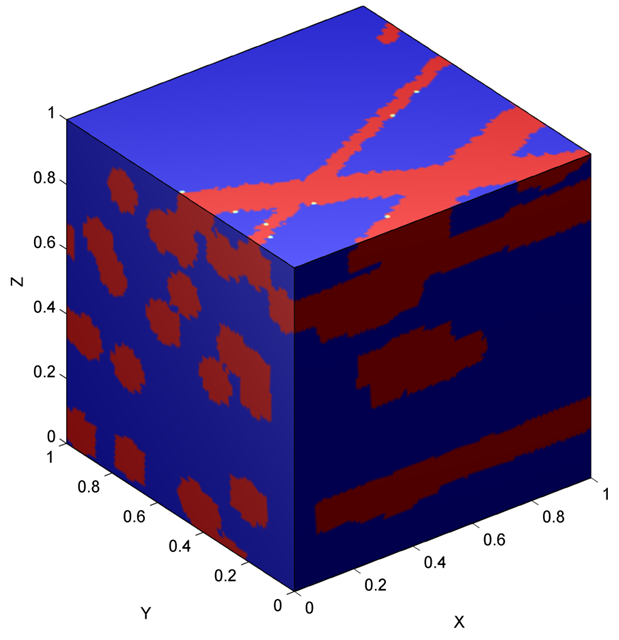}
  \includegraphics[width=.5\linewidth]{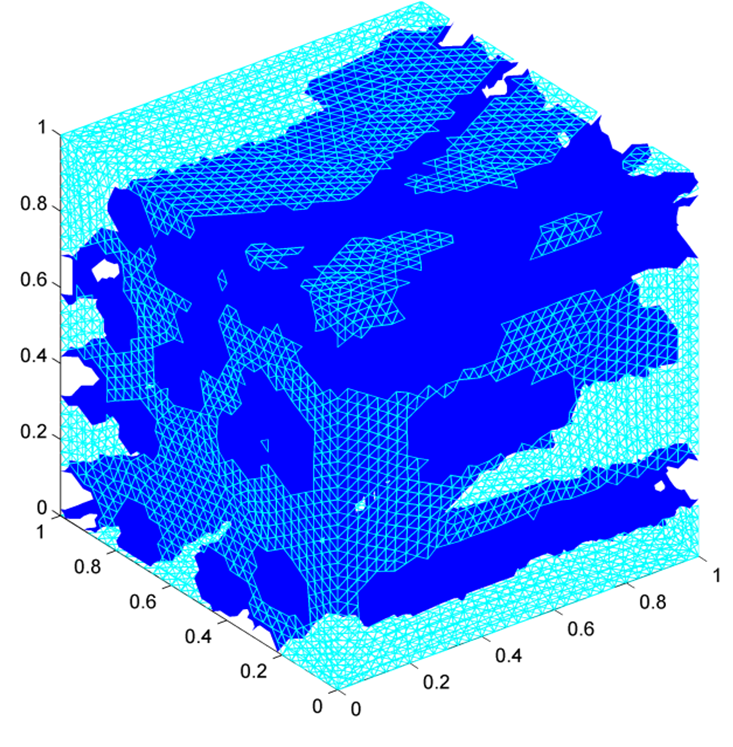}
  \caption{Computational domain imported to the finite element platform. The fibers are shown with red medium. The mesh used in the computation for estimating mechanical and electrical properties.}
  \label{fig:femimported}
\end{figure}

The generated three dimensional structure after the moving sphere method is illustrated in Fig.~\ref{fig:threeDmatlab}. Observe that the $yz$-plane at $x=0$, the cross section of the fibers are somewhat round due to the discretization used in generating the fibers. The fiber concentration (solid material) was adjusted with the number of lines in the line generating routine.
Once the structure in Fig.~\ref{fig:threeDmatlab} is generated, it was imported to the finite element software\cite{comsol43}. The imported computational domain and the mesh for the analysis are shown in Fig.~\ref{fig:femimported}. Similar approaches for electrical problems were used by the author previously to characterize electrical properties of clay filled polymers \cite{TuncerIPMHVC2010} and characterization of checkerboard structures \cite{TuncerCEIDP2012}.
The material properties for the fibers can be taken from the literature \cite{Pandini2011,Fukuhara2003}. 
Since we are interested in the change in elastic modulus, the absolute values play no significant role. However, the Poisson ratio is of importance and it was taken to be 0.40, see refences \cite{Pandini2011,Fukuhara2003}.

\begin{figure}[ht]
  \centering
  \includegraphics[width=\linewidth]{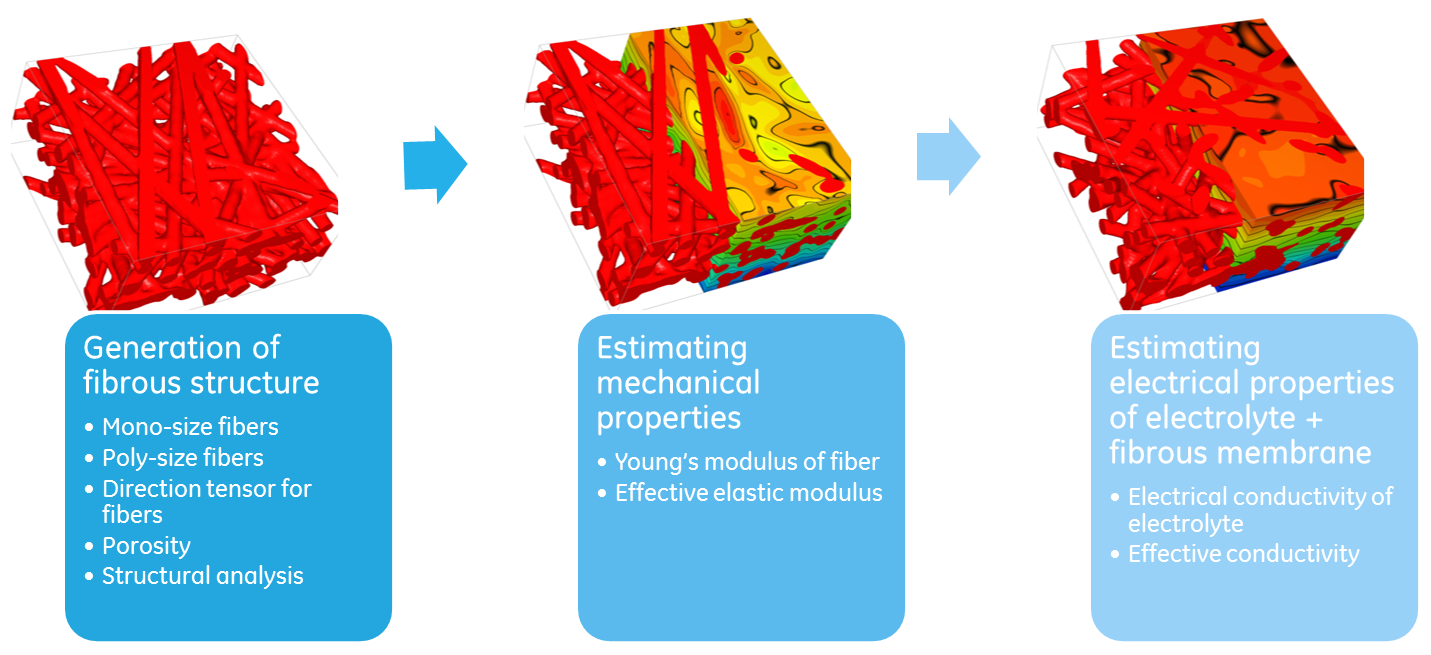}
  \caption{Structure of the computational model.}
  \label{fig:mdlstructure}
\end{figure}

In the electrical property simulations the fibers were considered to be perfect dielectrics and the surrounding medium is considered to be conducting electrolyte with conductivity $1$~S/m. The results were shown scaled to the electrolyte conductivity. The list of the steps the investigations are shown in Fig.~\ref{fig:mdlstructure}, the structures shown in the figure are taken from GeoDict \cite{GeoDict2005}.

Below we present the results from GeoDict and in-house-created program, respectively. It was hard to determine the geometrical description outputs of GeoDict with the in-house-created routines. The calculated effective properties were the only parameters to compare the both approaches. As stated before, we have also several experimental data to compare with the simulations.

\subsection{Modeling with GeoDict}
The GeoDict software is user-friendly and powerful, however it is time consuming to perform the Monte Carlo type simulations due to graphic-user-interface used to enter all parameters and read all output results. We studied 17 cases with different input parameters with two fiber fractions, 20\% and 30\% ($q$ in Table~\ref{tab:pbtGeoDict}). The estimated concentrations were very similar to those entered as an input--$q_c$ is the output volume fraction of fibers in Table~\ref{tab:pbtGeoDict}. The dimensions of the fiber radii $d$ were taken to be 5, and the computational domain $X_s\times Y_s\times Z_s$ were altered in the investigations to see the importance of finite size scaling. GeoDict estimates the fraction of fibers in three axes, which are shown in Table~\ref{tab:pbtGeoDict} with $f_x$, $f_y$ and $f_z$, respectively.
\begin{figure}[ht]
  \centering
  \includegraphics[width=\linewidth]{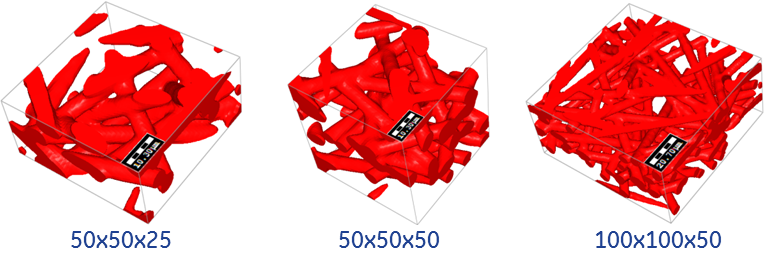}
  \caption{Finite size scaling in the computational domain, different sizes of domains considered in the simulations.}
\end{figure}
\begin{table}[ht]
  \caption{Simulation performed with GeoDict 
and various adjustable model parameters in GeoDict. The Young's moduli $E_{xx}$ and $E_{yy}$ are in MPa units and represent the elastic moduli values in $x$- and $y$-directions.}
  \centering
  \begin{tabular}{ccccccccccccc}
    \hline
    case&$q$&$q_c$&$d$&$X_s$&$Y_s$&$Z_s$&$f_x$&$f_y$&$f_z$&$E_{xx}$&$E_{yy}$&$\sigma_z^e/\sigma_0$\\
    \hline
    1&20&20.205&5&100&100&50&0.471&0.446&0.082&252&244&0.590\\
    2&20&20.091&5&100&100&50&0.453&0.477&0.070&115&267&0.596\\
    3&20&20.313&5&100&100&50&0.423&0.461&0.115&105&258&0.603\\
    4&20&20.111&5&100&100&50&0.430&0.475&0.095&112&248&0.597\\
    5&20&20.640&5&50&50&50&0.409&0.508&0.083&79&124&0.595\\
    6&20&20.592&5&50&50&25&0.415&0.445&0.140&190&89&0.614\\
    7&20&20.312&5&50&50&25&0.376&0.527&0.097&253&61&0.606\\
    8&20&20.658&5&50&50&25&0.377&0.526&0.098&202&128&0.548\\
    9&20&20.592&5&50&50&25&0.415&0.445&0.140&190&89&0.614\\
    10&20&20.126&5&50&50&25&0.519&0.398&0.082&219&108&0.595\\
    11&30&30.690&5&50&50&25&0.465&0.415&0.120&227&189&0.435\\
    12&30&30.150&5&50&50&25&0.527&0.355&0.117&190&121&0.455\\
    13&30&30.017&5&100&100&50&0.485&0.440&0.074&247&227&0.420\\
    14&30&30.176&5&100&100&50&0.428&0.457&0.115&210&252&0.435\\
    15&30&30.147&5&100&100&50&0.452&0.472&0.076&215&223&0.435\\
    16&30&30.051&5&100&100&50&0.439&0.441&0.120&229&233&0.431\\
    17&30&30.125&5&100&100&50&0.447&0.467&0.086&217&226&0.441\\
    \hline\hline
  \end{tabular}
  \label{tab:pbtGeoDict}
\end{table}
\begin{figure}[ht]
  \centering
  \includegraphics[width=\linewidth]{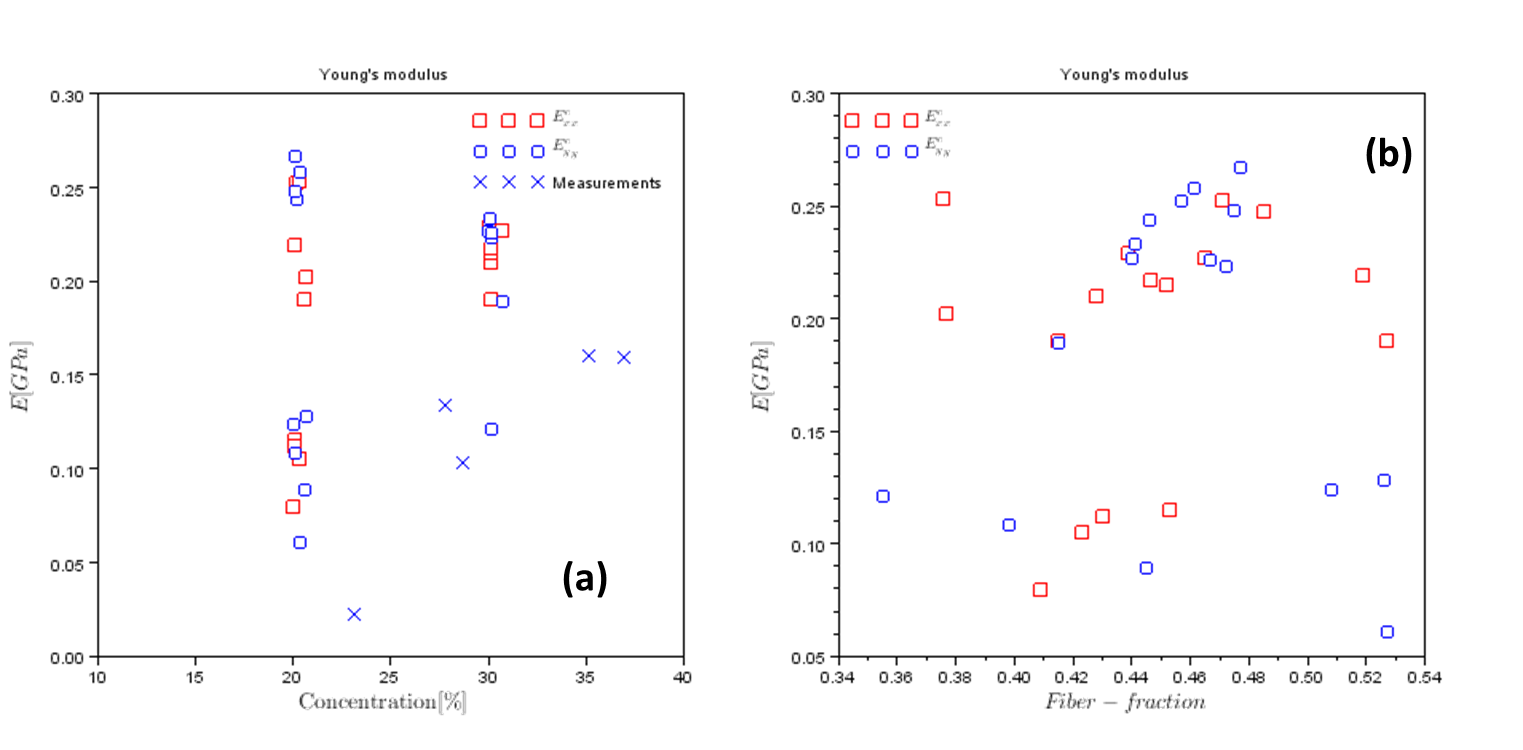}
  \caption{(a) Young moduli estimated with GeoDict and compared to experimental values for two different fiber concentrations. (b) Comparison of moduli data to fiber fraction in the estimated modulus direction.}
  \label{fig:Geodict1}
\end{figure}
The results from the GeoDict are listed in Table~\ref{tab:pbtGeoDict} for two different concentrations. First of all, no clear difference in the computational domain size was observed in the limited number of simulations. The main observation is that with increasing concentration of fibers, the spread of data was narrower. Compared to experimental values, which are shown in Fig.~\ref{fig:Geodict1}a with $\times$ symbols, the simulation values are overestimates, higher than the measurements. One reason for this can be the structures in the experimental samples, which were hardcore  systems. However, by post-processing with temperature and pressure one can convert these systems to softcore. To better understand the fraction of fibers in a given direction and the modulus in that direction, we plot the values of $f_i$ to $E_{ii}$, where $i={x,y}$. There is no clear dependency between these two parameters, as shown in Fig.~\ref{fig:Geodict1}b.

\begin{figure}[ht]
  \centering
  \includegraphics[width=.5\linewidth]{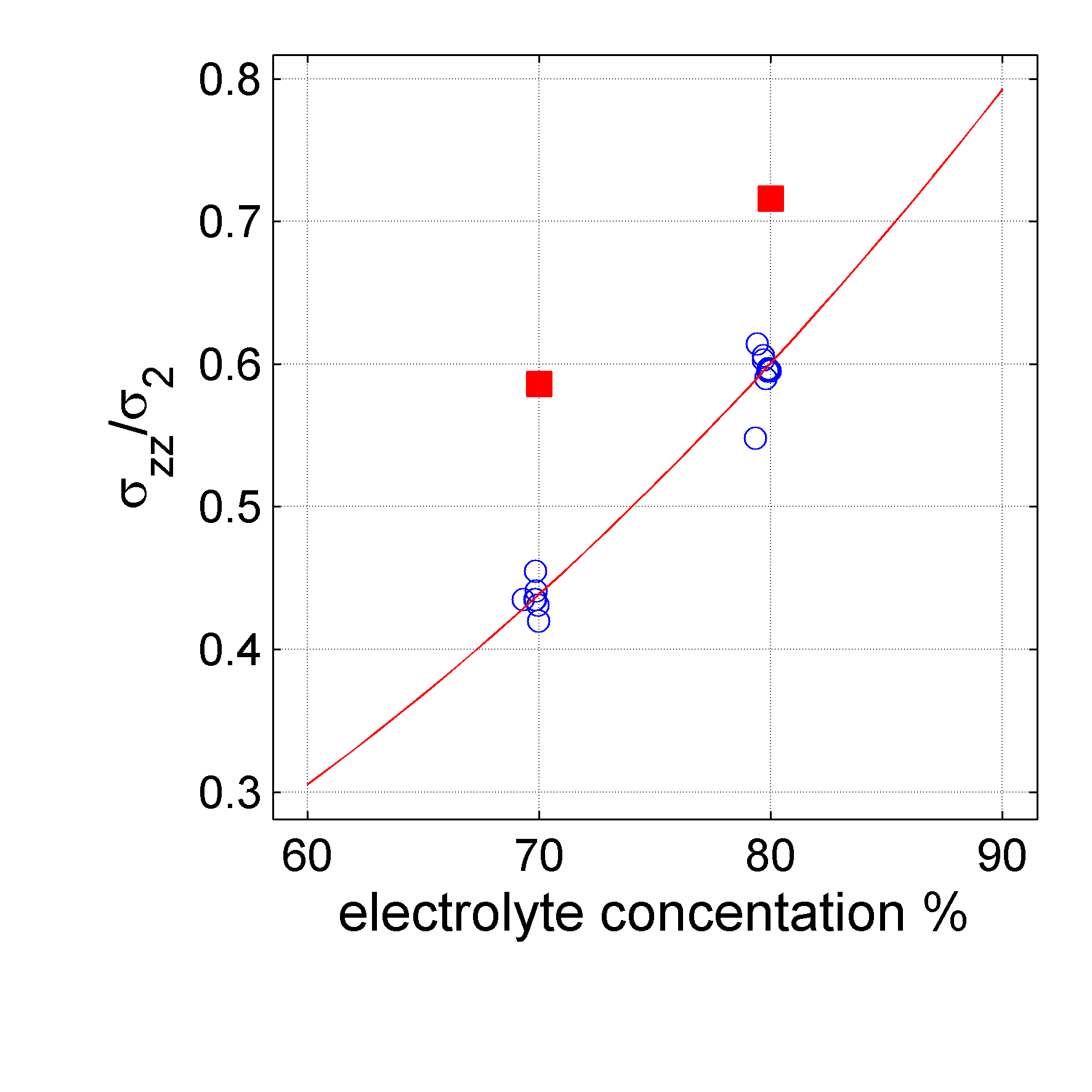}
  \caption{Electrical conductivities calculated with GeoDict. The solid squares ({\color{red}{$\blacksquare$}}) are the values estimated by using \citet{Arora2004}, and the solid line is from the Archie's law \cite{Achie} $\sigma_{zz}/\sigma_2=(1-q)^\alpha$ using $\alpha=2.35$.}
  \label{fig:GeodicCond}
\end{figure}
The electrical conductivities calculated with GeoDict are shown in Fig.~\ref{fig:GeodicCond}. The values for the given concentrations the Archie's law \cite{Achie}, $\sigma_{zz}=(1-q)^\alpha\sigma_2$ yields 0.716 and 0.586 with electrolyte conductivity $\sigma_2=1$ and $\alpha=1.5$, which was given by \citet{Arora2004}. Here we assume that the concentration of fibers are expressed with $q$. We have adopted the Archie's law and found that $\alpha=2.35$. The discrepancy in $\alpha$-values suggested by \citet{Arora2004} and our estimate is due to connectivity of the structures, the fibrous structures allow tortuous paths and have a high $\alpha$ value. For open pore structures with spheroidal pore sizes distributed over a range of shapes, one should expect low $\alpha$ values.

\subsection{Modeling with in-house routines}
\begin{figure}[ht]
  \centering
  \includegraphics[width=.5\linewidth]{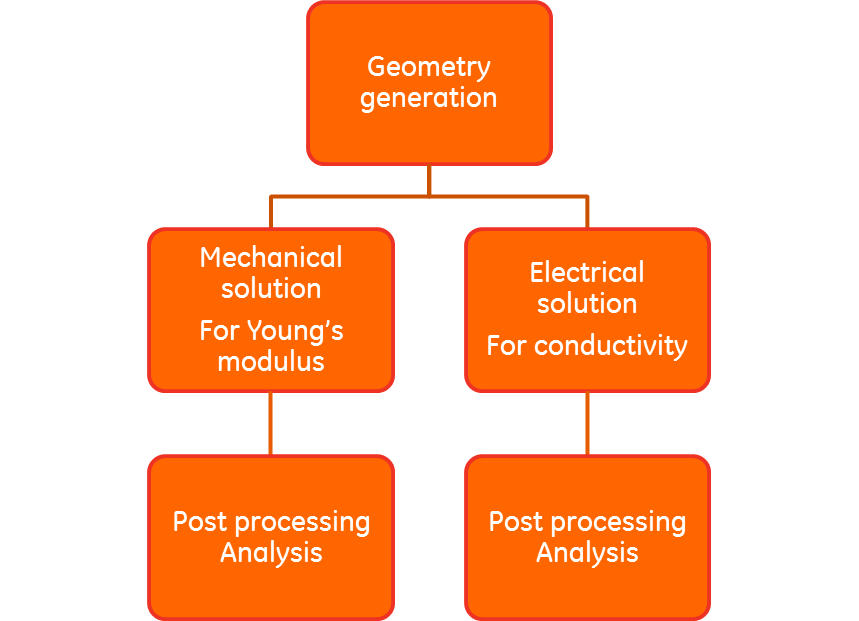}
  \caption{Procedures followed in the in-house developed routines.}
  \label{fig:inhouse}
\end{figure}
\begin{figure}[ht]
  \centering
  \includegraphics[width=\linewidth]{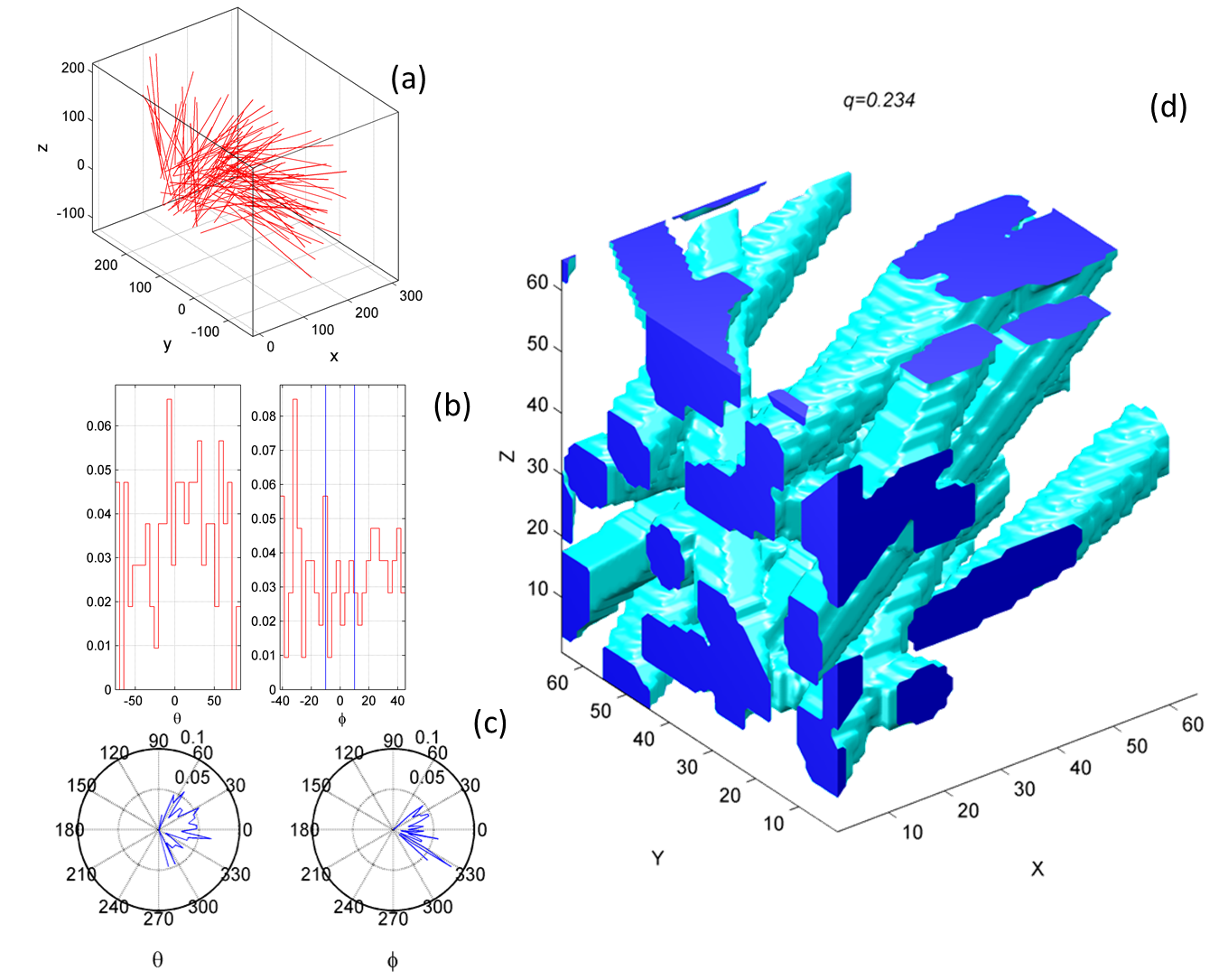}
  \caption{Additional example of a simulation pre-processor for the finite element method. (a) Generated infinite lines; (b) distribution of angles defined as in Fig.~\ref{fig:line}; (c) polar representation of the angles; (d) generated 3D structure using the lines in (a).}
  \label{fig:example2}
\end{figure}

The procedure structure in the simulations are shown in Fig.~\ref{fig:inhouse}. The top level procedure in Fig.~\ref{fig:inhouse} is the pre-processor and the middle levels are the finite element solver \cite{comsol43}. The bottom levels are the post-processing steps to determine the effective properties of the fibrous structures.  An additional example of pre-proccesing stage is presented in Fig.~\ref{fig:example2}.

Once the structure is defined as in Fig.~\ref{fig:example2}d, it is imported to the finite element platform and meshed as shown in Fig.~\ref{fig:femimported}. The meshing is an important part of the procedure and depending on the model, mechanical or electrical. Stored mesh structure requires significant computer memory, which is needed for solving the problem electrical and mechanical fields. Mechanical solutions required more memory due to the tensor nature of the material properties--both elastic modulus and Poisson's ratio needed to define anisotropic properties. Electrical problem assumed isotropic material properties.
The effective properties were estimated using averaging of physical parameters. 

\subsubsection{Mechanical calculations}
For mechanical approach the following procedure is followed; 
\begin{itemize}
\item $y$ and $z$ coordinate directions fixed with rolling boundary conditions at $y=0$ and $y=1$, and $z=0$ and $z=1$.
\item $x=0$ is fixed
\item $x=1$ a displacement is applied (uniform strain $S$).
\item Average modulus is estimated from $\bar{E^{xx}_e}=\int_{x=1} {\bf T_{x}}{\rm d}A/(A S)$, where ${\bf T_{xx}}$ is the stress in $x$-direction with strain applied in the $x$-direction, $S$ is the uniform strain. The parameter $A$ is the area of the surface at $x=1$, $A\equiv1$.
\item Material parameters are taken from the literature \cite{Pandini2011,Fukuhara2003}, 
and the void space is assume to have 1:1000 of the solid modulus with Poisson's ratio 0.33.
\item Simulation were performed with more than 1000 structures with varying concentration of fibers.	
\end{itemize}

\begin{figure}[ht]
  \centering
  \includegraphics[width=.6\linewidth]{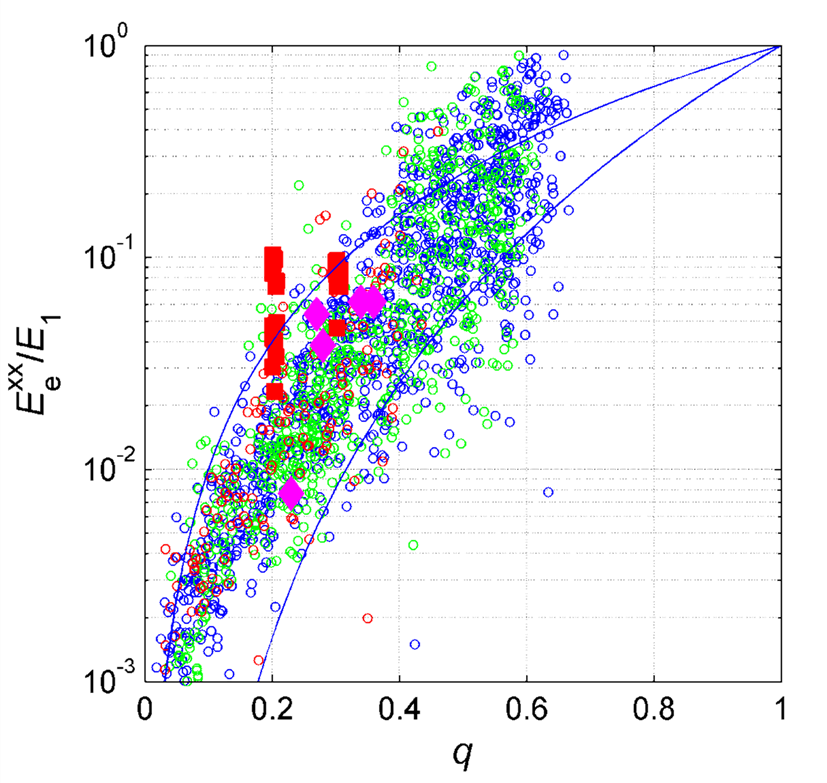}
  \caption{Simulation results from in-house developed routines for Young's modulus as a function of fiber concentration. The solid lines correspond to $q^2$ and $q^6$ relationships. The solid squares ({\color{red}{$\blacksquare$}}) are values from GeoDict and the solid diamonds are from experimental data.)}
  \label{fig:modulusinhouse}
\end{figure}
The results obtained for a large number of structures are shown in Fig.~\ref{fig:modulusinhouse}. We did not generated structures with over 60\% fiber volume fraction due to less interest in the application--the systems conductivity become very low for any potential application. We compare our simulations to those of the GeoDict results. However notice that to run an automated simulation with GeoDict to run $>1000$ simulations was not possible. We have overlaid the experimental data in Fig.~\ref{fig:modulusinhouse} as well. Different colors in open circles indicate different angular orientations in the figure. No significant influence of angle $\phi$ on mechanical properties observed for the considered low angle values ($\phi<30^o$). The experimental data were within the simulation results indicating that the model adopted was sufficient to generate realizations of the actual non-woven materials.

It is believed that due to finite size scaling the data has a large spread. For example a study performed by the author \cite{TuncerCEIDP2012} on checkerboards with sizes $16\times16$, $32\times32$ and $64\times64$ illustrated that with increasing computation domain size and increasing ratio between smallest and the largest units--for example in a $64\times64$ system, the ratio is 1/64--the spread in the distributions were becoming narrow. Similarly one can apply a similar rule here, and observe that the distributions would be narrow as the computation domain size increased. At the present we have no capability to perform these studies because of limitations in computation power.

\begin{figure}[ht]
  \centering
  \includegraphics[width=\linewidth]{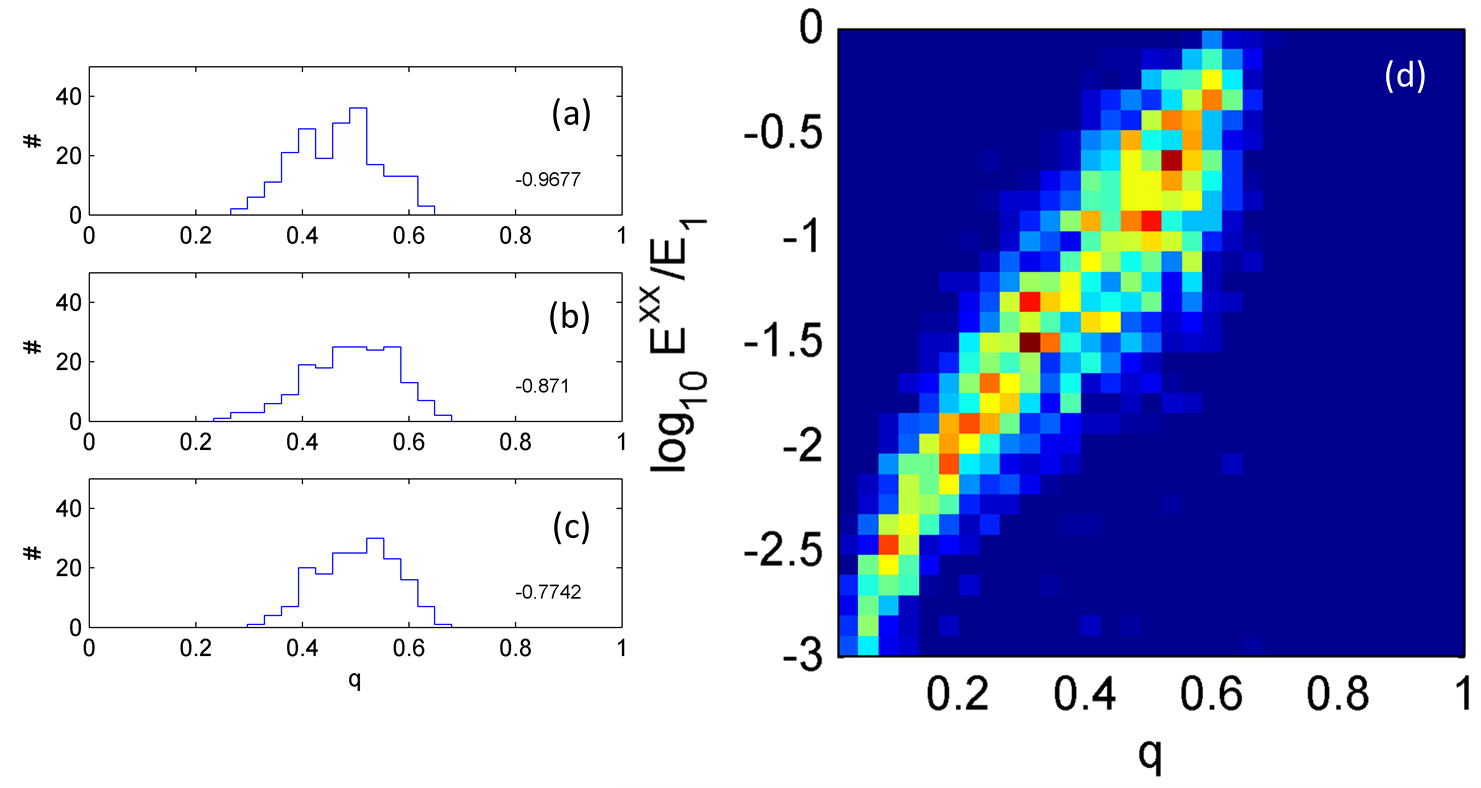}
  \caption{(a-c) One-dimensional histogram at different Young's modulus values. (d) Full two-dimensional histogram of the simulation results. The desired value of Young's modulus can be estimated from these figures.} 
  \label{fig:modulusinhouse2}
\end{figure}
To have a better understanding of the simulations, the data in Fig.~\ref{fig:modulusinhouse} is replotted in another representation. The data converted into a colored surface plot with light colors indicating the number density of the calculated values, a two-dimensional histogram. This is shown in Fig.~\ref{fig:modulusinhouse2}. The histograms indicate that the desired mechanical properties can be achieved with 0.40-0.55 volume fraction of fiber. It is recommanded that 350 MPa would be the desired  modulus for fabrication of membranes. In addition the value we report here for the fiber fraction region of interest is close to those separators available commercially as reported by \citet{Zhang2007351}. Notice that these results are obtained from the softcore system consideration.

\begin{figure}[ht]
  \centering
  \includegraphics[width=.6\linewidth]{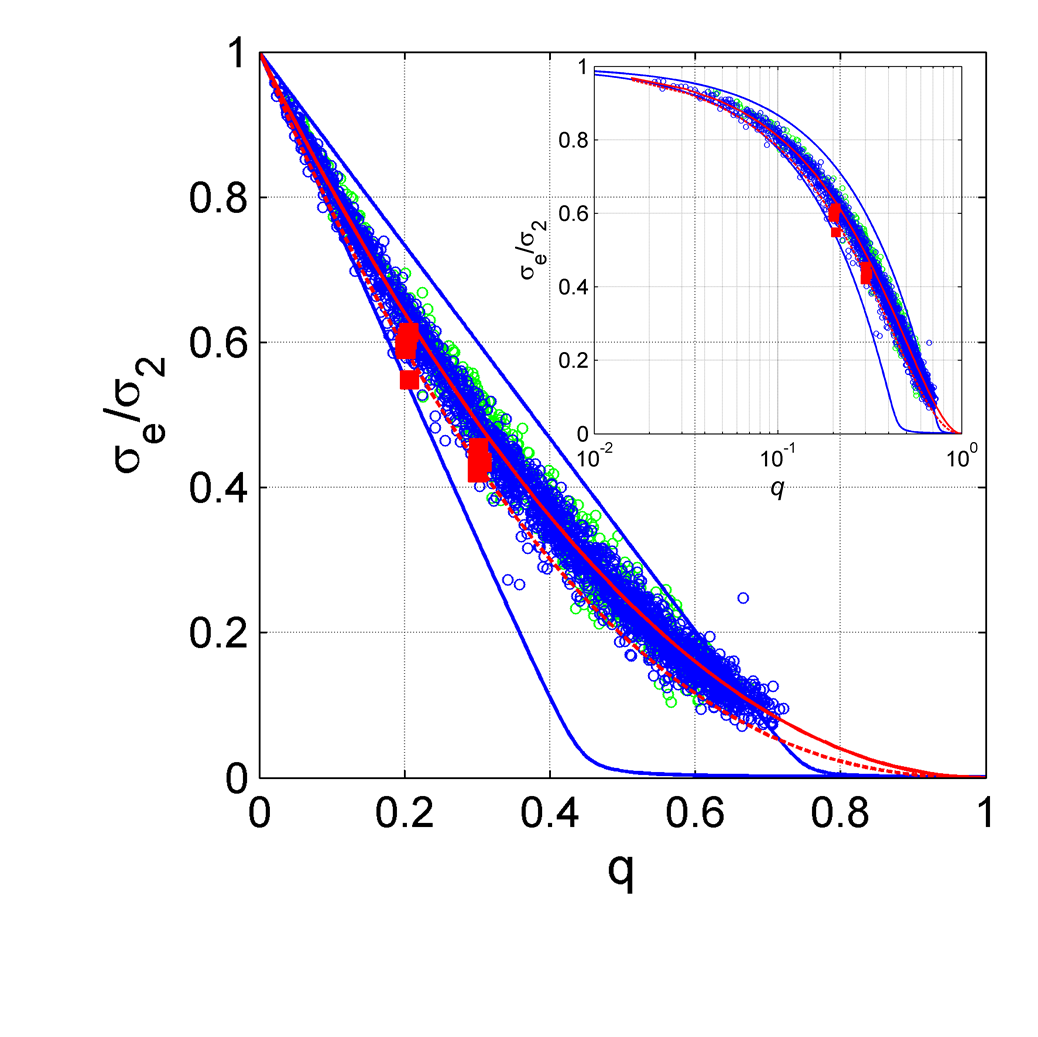}
  \caption{Effective conductivity of a fiber and electrolyte system as a function of fiber concentration. Simulation results  from in-house routines are shown with open symbols. Results from GeoDict are shown with solid squares ({\color{red}{$\blacksquare$}}). The solid lines that are used to define the bounds are estimated for particle filled composites using the Bruggeman expression \cite{Bruggeman1935} with shape parameters 1.8 and 4 for lower and upper bounds, respectively. The other lines that are drawn to represent the data are for Archie's law with $\alpha=2.35$ (estimated for GeoDict data) and $\alpha=2$ which represents the in-house results precisely. The inset is the same plot in logarithmic concentration axis. }
  \label{fig:condinhouse}
\end{figure}

\subsubsection{Electrical calculations}
The electrical effective properties are estimated as follows;
\begin{itemize}
\item $y$ and $x$ directions considered symmetrical boundary conditions.
\item $z=0$ is grounded, voltage is zero $V=0$.
\item $z=1$ voltage applied, $V=1$.
\item Conductivity is then estimated from the average current on $z=1$ surface, $\sigma_{zz}=\int J{\rm d}A/(V/t)$, where $J$ is the current, $A$ is the area, $V$ is the voltage and $t$ is the thickness of the sample, $t=1$.
\item Simulation were performed with more than 1000 structures.	
\end{itemize}
There is a clear trend of decreasing conductivity with increasing fiber amount--less electrolyte in the system as we increase the fiber amount--as shown in Fig.~\ref{fig:condinhouse}. Again the simulations from Geodict are shown in Fig.~\ref{fig:condinhouse}, In the figure data of the developed code (open circles) have similar results as those from GeoDict. Different colors in open circles indicate different angular orientations. No significant influence of the orientation of fibers in the effective conductivity. The solid lines in Fig.~\ref{fig:condinhouse} are trends estimated for different shapes, (rods and disks) using the Bruggeman symmetrical expression \cite{Bruggeman1935}. The shapes are obtained using 1.8 and 4 in Bruggeman expression. The shapes correspond to deformed spheres, oblate and prolate spheroids, indicating that the porous structure composite such pores. 

It is striking that the data obtained from the in-house routines can be expressed with Archie's law with exponent $\alpha=2$. This can be explained with using Bruggman's approach as well such that electrolyte is a continuous phase and starts to percolate even at very high concenration of fibers, notice that in our assumptions we have used low angles of $\phi$ in generating structures due to the orientation of fibers spinned in the experiments. There would not be a metal-insulator transition due to the connectivity of electrolyte at some fiber concentration. However, in a pure 2D problem, where fibers are uni-directional, the fibers would create states of blocking conduction at concenrations over 0.5 \cite{Percolation} with exponent 2. Since our first assumption imposes percolation at $q=1$, and low $\phi$ values yield quasi-2D, the percolation exponent of 2 could be used to describe the electrolyte conduction in fibrous non-woven structures.

\section{Conclusion}
An attempt to model non-woven fibrous mats for electrical energy storage devices was performed with numerical tools. An in-house developed numerical approach was employed, and its results were compared to a commercially available program and experimental results. It was shown that the current tools were capable of explaining the spread in the experimental data. It was found that fiber volume fractions about 0.40-0.55 would be of interest in the material fabrication because of required mechanical properties. The electrical properties of the electrolyte-fiber structure show a quadratic relation on electrolyte volume fraction.

\subsection*{Acknowledgements}
The project was supported by SABIC Innovative Plastics. The data and the scanning electron microscopy images were provided by Huiqing Wu, Yanju Wang, Jie Gao and Qunjian Huang from General Electric Corporate Technology Center in Shanghai. We appreciated their assistance. I would also like to express my thanks to Andreas Wieg mann from Fraunhofer Institute in Germany for allowing me to test the demo version of GeoDict for a month in early 2012.

\end{document}